\documentclass[11pt]{article}

\usepackage[preprint]{acl}

\usepackage{times}
\usepackage{latexsym}

\usepackage[T1]{fontenc}

\usepackage[utf8]{inputenc}

\usepackage{microtype}

\usepackage{inconsolata}

\usepackage{graphicx}

\title{Explainable Progressive Ancient Manuscript Duplicates Finder}
\title{Explainable Coarse-to-Fine Ancient Manuscript Duplicates Finder}
\title{Explainable Coarse-to-Fine Ancient Manuscript Duplicates Discovery}

\author{Chongsheng Zhang, Shuwen Wu, Yingqi Chen, Yi Men, Gaojuan Fan\\
  Henan University, China \\
  \texttt{\{Cszhang,ShuwenWu,Wingkeichanchan,Yi.Men,Fangaojuan\}@henu.edu.cn} \\ 
 \AND Matthias A\ss{}enmacher, Christian Heumann \\
  LMU Munich, Germany\\
\texttt{\{matthias,chris\}@stat.uni-muenchen.de} \\\And Jo{\~{a}}o Gama \\
University of Porto, Portugal\\
\texttt{jgama@fep.up.pt}}

\begin{document}
\maketitle
\begin{abstract}
Ancient manuscripts are the primary source of ancient linguistic corpora. However, many ancient manuscripts exhibit duplications due to unintentional repeated publication or deliberate forgery. The Dead Sea Scrolls, for example, include counterfeit fragments, whereas Oracle Bones (OB)  contain both republished materials and fabricated specimens. Identifying  ancient manuscript duplicates is of great significance for both archaeological curation and ancient history study.  In this work, we design a progressive OB  duplicate discovery framework that combines unsupervised low-level keypoints matching with high-level text-centric content-based matching to refine and rank the candidate OB duplicates with semantic awareness and interpretability. We compare our model with state-of-the-art content-based image retrieval and image matching methods, showing that our model  yields comparable recall performance and the highest simplified mean reciprocal rank scores for both Top-5 and Top-15 retrieval results, and with significantly accelerated computation efficiency. We have discovered over 60 pairs of new OB duplicates in real-world deployment, which were missed by domain experts for decades. Code, model and real-world results are available at: \url{https://github.com/cszhangLMU/OBD-Finder/}.
\end{abstract}

\section{Introduction}
Ancient manuscripts are the key source for ancient language corpora. However, many ancient manuscripts contain duplicates, due to unintentional repeated publication or deliberate forgery. For instance, the Dead Sea Scrolls contain forged fragments \cite{dssforgeries}, while Oracle Bones (OB) contain both repeated publications and forged ones. Finding  ancient manuscript duplicates  can help identify forgeries, eliminate duplicate fragments and prevent redundant research, while offering the potential to correct erroneous fragment rejoinings. Moreover, it facilitates empirical study on the damage and deterioration of ancient manuscripts  during their circulation.

In particular, the identification of Oracle Bone  duplicates has been a fundamental research issue in Oracle Bone Inscription (OBI) research. OBI was used in the late Shang Dynasty more than 3000 years ago for divination and recording purposes. But from then on, these Oracle Bones had been buried underground for thousands of years, until they were rediscovered in the year of 1899 for containing inscribed ancient Chinese characters. Due to drilling and burning before and during divination, and the long-term underground corrosion, as well as excavation, transportation, and circulation after their excavation, about 90\% of the OBs have been fragmented and are now scattered in different collections around the world \cite{obrejoin}.

\begin{figure*}[ht]
\centering
\includegraphics[width = \linewidth]
{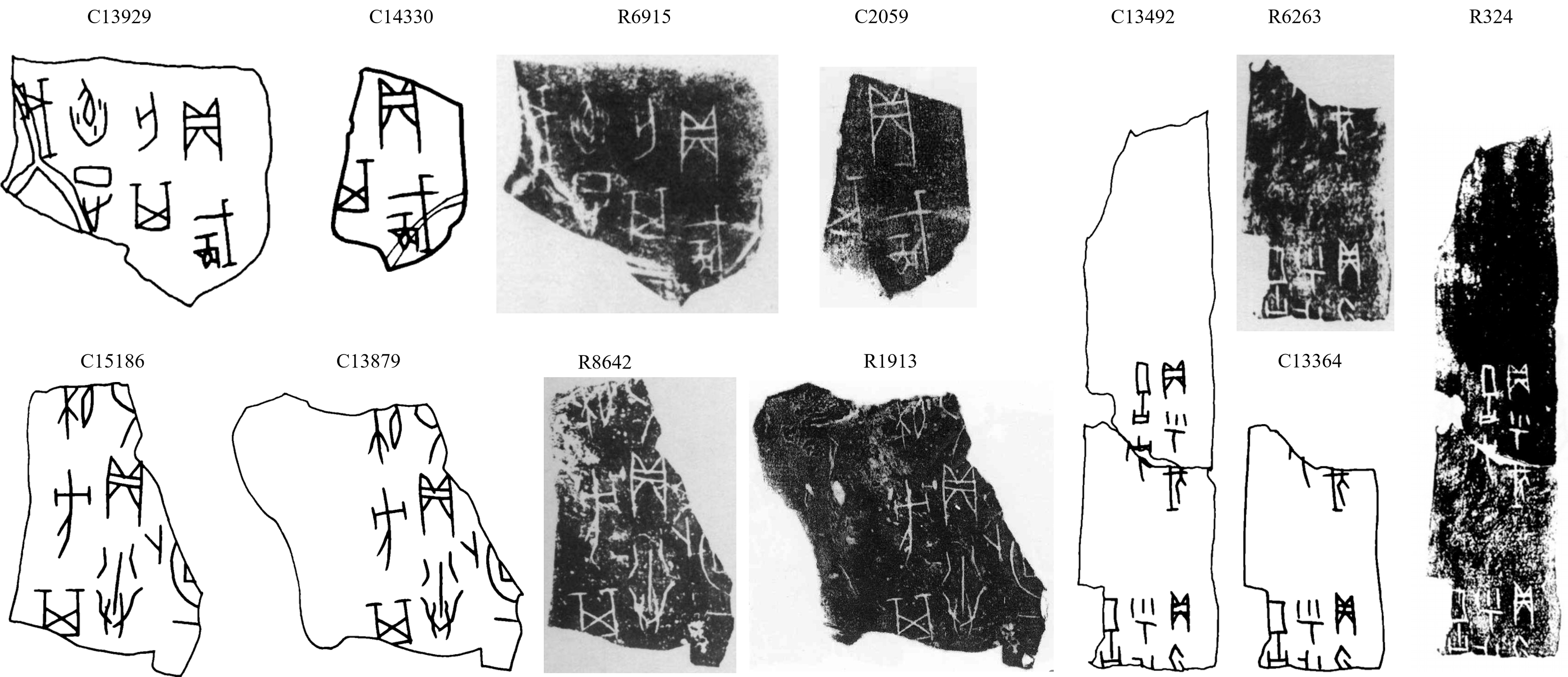}
\caption{Three groups of new Oracle Bone duplicates discovered by our model, which have been missed by domain experts for decades. For each group of duplicate, we provide both the manual copies and rubbings of the Oracle Bones. We can see that, finding Oracle Bone duplicate is similarity-based matching, rather than exact matching, and there exists both one-to-one (e.g.the bottom left pair) and one-to-many matchings (e.g.the top left pair and the right pair). Note that, in our implementation, we only use the manual copies of the Oracle Bones.}
\label{fig:0}
\end{figure*}

As precious cultural relics, many Oracle Bones were circulated among various collectors and antique dealers in the initial period after their discovery in 1899. Limited by communication and dissemination methods at that time, the same OBs might have been repeatedly published in different publications at different times in different locations, which led to the phenomenon of OB duplicates, denoting that  the fragments were repeatedly published. Some OBs  further fragmented during circulation; on the other hand, as OBI research advances, some fragmentary OBs might have been rejoined by OBI domain experts and republished again (e.g., the right group of duplicates in Figure \ref{fig:0}). As such, OB duplicates exhibit both one-to-one and one-to-many image matching relationships. Although domain experts have manually found many duplicates in their research, given the huge cardinality of OB fragments (more than 160,000), AI-enabled OB duplicates discovery becomes imperative.

Oracle Bone Inscription is carved writing, its main research materials and  publication formats are rubbings and manual copies. For  rubbing materials,  people place papers onto the surface of the Oracle Bones, then use Rubbing (with inks) to copy the carved inscriptions. Domain experts can also reproduce (copy) the carved inscriptions by hand, which is named Manual Oracle Bone Inscriptions Copy, referred to as OB manual copies for short. Figure \ref{fig:0} presents examples for both formats, which are actually cases of the new OB duplicates discovered by our model. Comparing the two formats,  manual OBI copies rely on domain knowledge but has no background noises, whereas OBI rubbings often contains substantial noise disturbance, although domain knowledge is not required. Both formats can keep the original sizes of the Oracle Bones, which is not possible when using cameras.  In 2022, the largest collection of manual OBI copies was published  \cite{huangobicopybook}, for which a large team of OBI researchers invested 10 years to create high-quality manual OBI copies for around 60,000 OBs.

With this new collection at hand, in this work we aim to devise a comprehensive framework for discovering OB duplicates at large-scale. Since different domain experts have slightly different copying styles for the same OBIs (such as variations in pen movement, stroke thickness, brush pressure), finding OB duplicates can be essentially formulated as a content-based image retrieval (CBIR) \cite{Smooth-AP,Proxy-Anchor,HybridHash} or image matching \cite{LoFTR,Omniglue,MINIMA} task.

\textit{\textbf{Contributions}}. To our knowledge, this work is among the first technical efforts that investigate AI-enabled Oracle Bone duplicates discovery. We design OBD-Finder, an explainable coarse-to-fine text-centric Oracle Bone duplicates discovery framework that successively utilizes unsupervised low-level key feature points matching and high-level content/character similarity for ranking the OB duplicate candidates. We have deployed our model in real-world applications, where we have successfully identified 63 pairs of new Oracle Bone duplicates, which have been verified by OBI community. Figure \ref{fig:0} presents three groups of new OB duplicates discovered by our model.

We also conduct extensive experiments on a large dataset of OB copies from \cite{huangobicopybook}.  We compare our model with state-of-the-art CBIR and image matching methods, showing that our model achieves Top-K recall performance comparable to state-of-the-art methods, but with significantly accelerated computational efficiency and substantially reduced GPU memory consumption.  Our model also attains the highest simplified mean reciprocal rank scores for both Top-5 and Top-15 retrieval results, demonstrating that it excels at prioritizing correct matches. 

\section{Related Work}
AI-enabled ancient manuscript fragments retrieval and rejoining have very important real-world applications \cite{fragretrieval,obrejoin,Dunhuangrestore}. Our  domain-specific  Oracle Bone (OB) duplicates discovery problem can be essentially formulated  as a image retrieval or image matching task. In the following, we provide a brief overview of the content-based image retrieval (CBIR) and image matching techniques assessed in this study.


\paragraph{Content-based image retrieval.} The Smooth-AP loss \cite{Smooth-AP} provides a differentiable approximation for Average Precision  to enable end-to-end training for ranking-based CBIR tasks. HashNet \cite{HashNet} adopts a sign activation function  for binarizing the K-dimensional deep feature representation into K-bit binary hash code for retrieval tasks. In this paper, we use latest transformer-based implementation of HashNet \cite{ViTHash}. HybridHash \cite{HybridHash} is a hybrid deep hashing architecture combining self-attention and convolutional layers for enhancing image retrieval performance. Note that, since there also exists one-to-many mapping relationship in OB duplicates, contour-based retrieval methods are not applicable in our task. 

\paragraph{Image matching.}  Image matching, also known as feature/keypoint matching, aims to establish correspondences between points in two images depicting the same scene or object. SIFT (Scale-Invariant Feature Transform) \cite{SIFT} is a classic algorithm for image matching. In recent years, deep learning based techniques have substantially improved the state-of-the-art in this field. LoFTR \cite{LoFTR} utilizes Transformer self-attention and cross-attention mechanisms to establish  coarse-level matches of the keypoints, followed by fine-level match on the cropped local windows for each coarse match. OmniGlue \cite{Omniglue} leverages vision foundation models for boosting generalization to unseen domains. It also uses self- and cross-attention for establishing intra- and inter-image connectivity graphs to enhance correspondence estimation. MINIMA \cite{MINIMA} is a latest image matching technique that  contains a simple data engine for freely generating multiple modalities for images to pre-train modality invariant image matching model, achieving state-of-the-art performance. 

Despite their technical advancement and substantial performance improvement, OmniGlue essentially depends on foundation models for accurate image matching, while MINIMA relies on ``heavy'' pre-training on generated multi-modal data, and both of them have significantly slower inference speed, compared to our proposed framework.  Moreover, all the above image matching  methods depend on low-level feature matching, but lack high-level semantic guidance. In comparison, in this work we seamlessly integrate unsupervised low-level feature matching and high-level character-centric semantic-aware content matching, offering outstanding retrieval and matching performance, fast inference speed, and strong interpretability  for OB duplicates discovery.   

\section{Methodology}

\subsection{Framework}
As can be seen from Figure \ref{fig1}, we propose a progressive coarse-to-fine Oracle Bone duplicate discovery framework, namely OBD-Finder, which combines unsupervised low-level keypoint matching with high-level, character-centric content-based image matching. Keypoint matching operates at low-level visual feature scale, which can prune out candidates with low degree of match in the initial stage, but lacks explicit semantic supervision and interpretability. Our framework bridges this gap by first grouping the keypoints based on  their association with the character regions, then assesses the global matching degree between the two groups of keypoints via character-level visual content similarity computation. This dual matching mechanism enhances Oracle Bone duplicates discovery accuracy through a progressive coarse-to-fine refinement manner, by effectively and seamlessly integrating both low-level keypoint and high-level character-based semantic cues, resulting in  more accurate and semantic-aware image matching. Our framework consists of four  subsequent steps: 

\begin{figure*}[ht]
\centering
\includegraphics[width=0.9\textwidth]{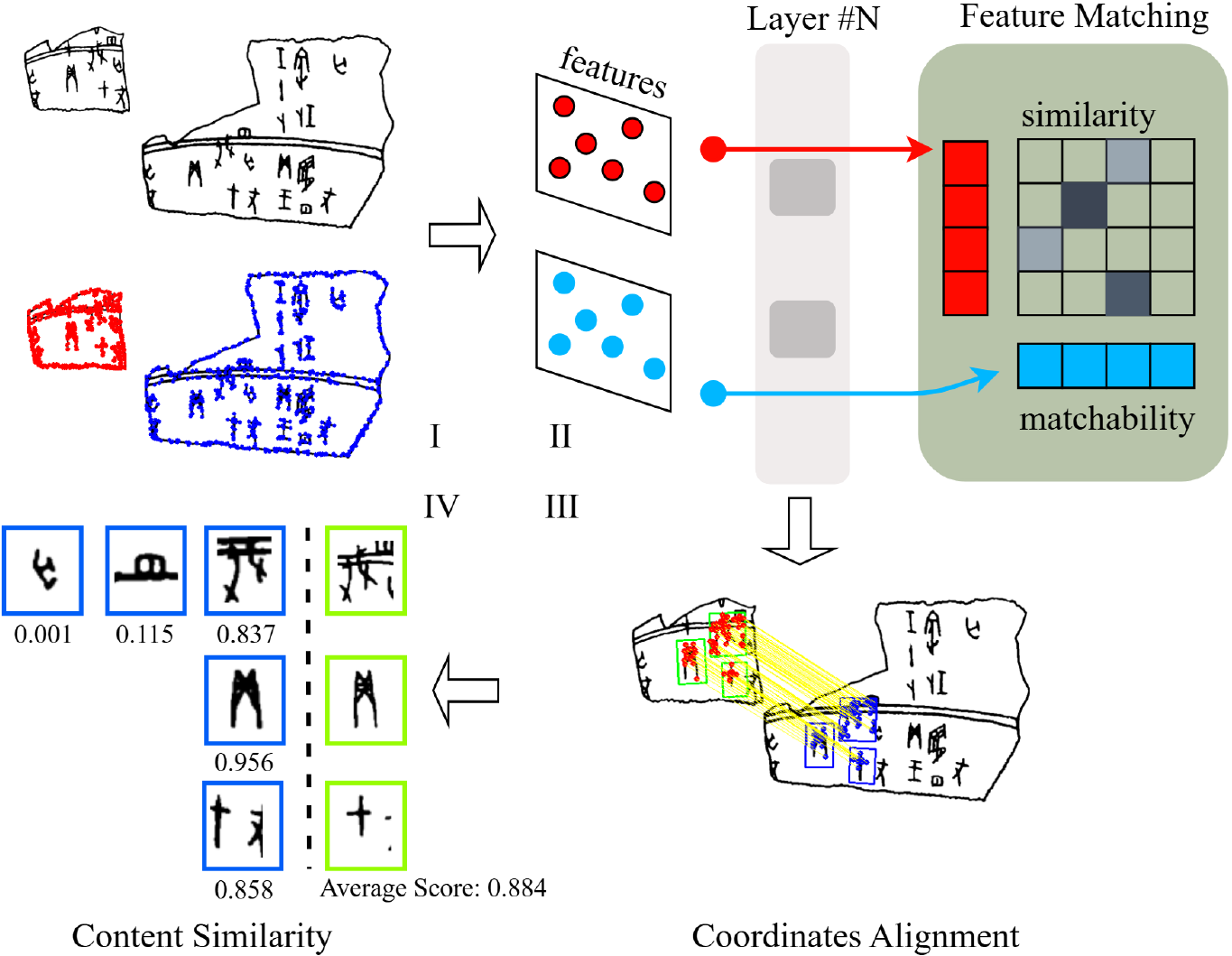}
\caption{The overall framework of OBD-Finder for Oracle Bone duplicates discovery. } \label{fig1}
\end{figure*}

\begin{enumerate}
    \item \textbf{Feature Extraction.} We perform unsupervised  keypoints extraction on the OBs using a pre-trained  model \cite{SuperPoints}.
    \item \textbf{Feature Matching.} We next  apply unsupervised keypoints mapping between the two OB images using a pre-trained model  \cite{LightGlue}. Candidate with low overall matching degrees will be filtered out. 
    \item \textbf{Coordinate Alignment.} After obtaining the correspondence between the keypoints in feature matching, we apply affine transformations for each image pair, in which we map the coordinates of the image with fewer feature points to  the other image. 
    \item \textbf{Character-level Content Similarity.} we then localize the Oracle Bone characters in each image, using a  text detector  \cite{EAST}. Given that the coordinate systems of two images are aligned,  for each character in the smaller image,  we search for the overlapped characters in the counterpart image, next compute the content similarity between them, using a simple Siamese network model \cite{DeepFace}. 
\end{enumerate}

\subsection{Key Features}

OBD-Finder for ancient manuscript duplicates discovery has four primary characteristics:

\begin{enumerate}
    \item  It is a progressive coarse-to-fine framework that seamlessly proceeds from low-level keypoints matching to high-level semantic-aware content similarity computation, resulting in very accurate OB duplicates discovery.  
    
   \item  It is a transparent framework with strong interpretability. 
   
   \item It is unsupervised and almost training-free, only requiring little annotation effort. 
   
   \item   It is highly efficient, compared to state-of-the-art image matching methods, which we will demonstrate in the empirical studies. 
\end{enumerate}

\section{Experiments}
\begin{table*}[htbp]
\caption{Comparing OBD-Finder with CBIR methods (Recall@K performance).}
\label{tab:1}
\centering
\begin{tabular}{l|ccccc}
\hline
Method & Recall@1 & Recall@5 & Recall@10 & Recall@15 & Recall@20 \\
\hline
Smooth-AP \cite{Smooth-AP} & 34.1 & 62.9 & 73.8 & 78.8 & 82.8 \\
Proxy-Anchor \cite{Proxy-Anchor} & 71.5 & 77.0 & 84.1 & 86.6 & 91.9 \\
HashNet \cite{ViTHash} & 57.6 & 63.8 & 69.3 & 78.4 & 85.8 \\
HybridHash \cite{HybridHash} & 60.8 & 66.8 & 73.2 & 79.3 & 88.3 \\
Ours (OBD-Finder) & \textbf{80.0} & \textbf{85.3} & \textbf{90.4} & \textbf{94.3} & \textbf{98.0} \\
\hline
\end{tabular}
\end{table*}

\begin{table*}[htbp]
\caption{Comparing OBD-Finder with image matching methods (Recall@K).}
\label{tab:2}
\centering
\begin{tabular}{l|cccccc}
\hline
Method & Recall@1 & Recall@5 & Recall@10 & Recall@15 & Recall@20 & Recall@25 \\
\hline
SIFT \cite{SIFT} & 33.3 & 40.0 & 46.8 & 53.3 & 66.6 & 73.2 \\
LoFTR \cite{LoFTR} & 73.6 & 75.4 & 81.3 & 86.0 & 92.3 & 98.3 \\
OmniGlue \cite{Omniglue} & 82.5 & 86.0 & 91.2 & 95.6 & 98.2 & \textbf{100} \\
MINIMA \cite{MINIMA} & \textbf{84.4} & \textbf{90.5} & \textbf{94.7} & \textbf{98.2} & \textbf{100} & \textbf{100} \\
Ours (OBD-Finder) & 80 & 85.3 & 90.4 & 94.3 & 98 & \textbf{100} \\
\hline
\end{tabular}
\end{table*}

\subsection{Experimental Setup}
\paragraph{\textbf{Dataset.}} We use the 60,000 manual Oracle Bone copies from the Oracle Bone Inscription Copy Series \cite{huangobicopybook} as the retrieval database, which have been divided into 25 OB groups by the OBI community based on dating and writing style. Two OBI domain experts who are co-authors of this work (Prof. Yi Men and Ms. Yingqi Chen) have manually collected 150 pairs of OB duplicates published in the OBI literature, which belong to different OB groups and are used the query images (with ground-truths). For each pair of images, we perform OB duplicate checking in the corresponding OBI group. For OBI character localization using EAST \cite{EAST} and character-level content matching using Siamese network \cite{DeepFace},  OBI domain experts have manually annotated the character regions and categorization for 500 OBs.   

\paragraph{\textbf{Evaluation Metrics.}} We use two evaluation metrics: Recall@K and a simplified mean reciprocal rank measure MRR@K. Recall@K measures how many relevant items were successfully retrieved in the Top-K results, but doesn't consider their rankings. To address this limitation, we also adopt a modified version of MRR@K, since our dataset only contains one correct match per query, we simply average the rank position of the correct answer  in the Top-K results across all query images,  abbreviated as Rank@K.

\subsection{Results}
We first compare our model with content-based image retrieval (CBIR) methods. As reported in Table \ref{tab:1}, OBD-Finder consistently obtains the best  recall performance than the CBIR methods. 

\begin{figure*}[htbp]
\includegraphics[width=\textwidth]{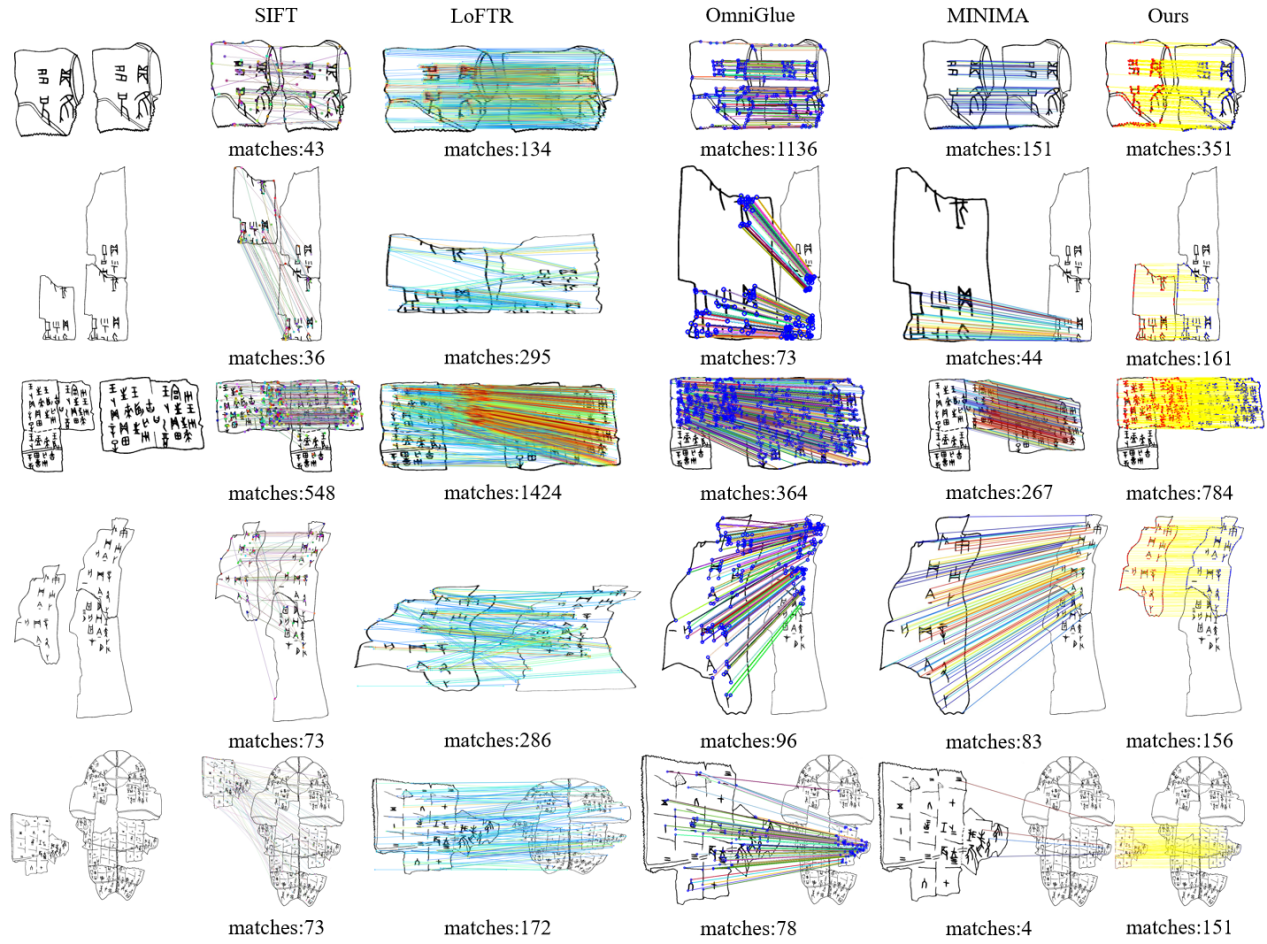}
\caption{Case studies for image matching results of different methods.} \label{fig2}
\end{figure*}

We next compare our model  with state-of-the-art image matching techniques. As shown in  Table \ref{tab:2}, OBD-Finder exhibits recall performance on par with latest image matching methods. In Figure \ref{fig2}, we showcase image matching results of different methods. Moreover, as reported in Tables \ref{tab:3}, our model obtains the best Rank@K scores in the Top-5 and Top-15 retrieval results, demonstrating that it excels at prioritizing correct matches.

\begin{table*}[htbp]
\caption{Comparing OBD-Finder with image matching methods (Rank@K).}
\label{tab:3}
\centering
\begin{tabular}{l|ccccc}
\hline
Method & Rank@5 & Rank@10 & Rank@15 & Rank@20 & Rank@25 \\
\hline
SIFT \cite{SIFT} & 1.6 & 2.70 & 4.16 & 6.15 & 7.74 \\
LoFTR \cite{LoFTR} & 1.09 & 1.88 & 2.74 & 3.76 & 5.22 \\
OmniGlue \cite{Omniglue} & 1.13 & \textbf{1.51} & 2.05 & 2.52 & 2.89 \\
MINIMA \cite{MINIMA} & 1.36 & 1.66 & 2.07 & \textbf{2.38} & \textbf{2.38} \\
Ours (OBD-Finder) & \textbf{1.06} & 1.63 & \textbf{2.00} & 2.61 & 2.93 \\
\hline
\end{tabular}
\end{table*}

\begin{table*}[htbp]
\caption{Comparisons on Inference speed, FPS, and GPU memory consumption.}
\label{tab:4}
\centering
\small 
\setlength{\tabcolsep}{4pt} 
\begin{tabular}{l|cccc}
\hline
Method & Recall@20 & Inf. Speed (s/pair) & FPS (pair/s) & GPU (MiB) \\
\hline
SIFT \cite{SIFT} & 66.6 & 0.017 & 59 & N/A \\
LoFTR \cite{LoFTR} & 92.3 & 3.6 & 0.28 & 3502 \\
OmniGlue \cite{Omniglue} & 98.2 & 45 & 0.02 & 23612 \\
MINIMA \cite{MINIMA} & 100 & 1 & 1 & 14890 \\
Ours (OBD-Finder) & 98 & 0.021 & 47.62 & 5215 \\
\hline
\end{tabular}
\end{table*}

In Tables \ref{tab:4},  we  observe that OBD-Finder has  significantly faster inference speed than state-of-the-art image matching algorithms (40 times faster), but with substantially less GPU consumption (1/3 of their GPU usages). Therefore, our model is accurate and efficient at prioritizing correct matches.

\subsection{Real-world Deployment}
\begin{figure*}[ht]
\centering
\includegraphics[width = \linewidth]
{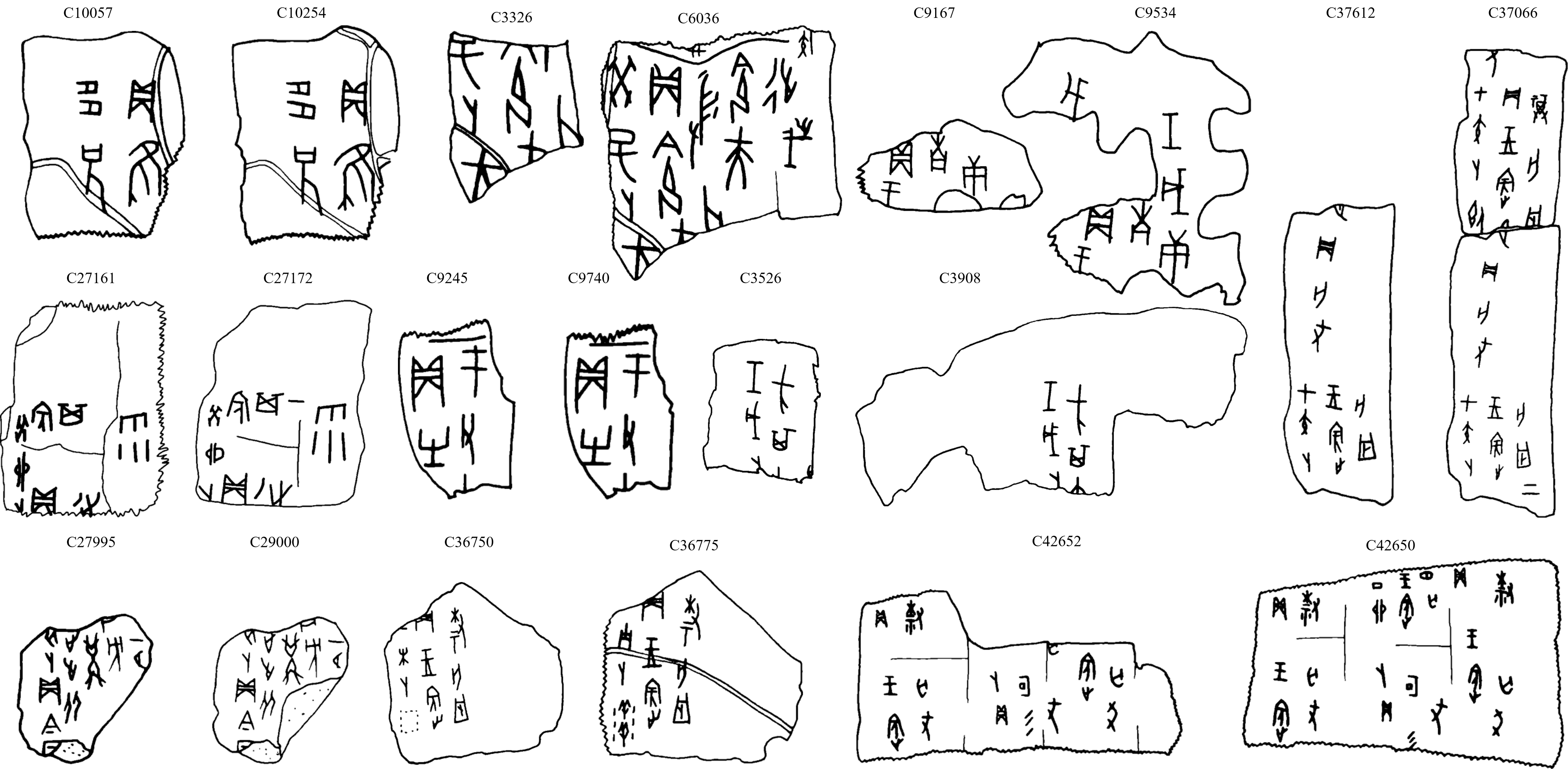}
\caption{Ten groups of new Oracle Bone duplicates discovered by our OBD-Finder framework.}
\label{fig:10}
\end{figure*}

Besides empirical study, we also perform OB duplicate discovery for other OBs in each category, where we have successfully discovered 63 groups of new OB duplicates, which have been verified by OBI domain experts (Prof. Yi Men and Ms. Yingqi Chen, who are also co-authors of this work). Besides Figure \ref{fig:0}, in Figure \ref{fig:10}, we showcase ten more  pairs of new Oracle Bone duplicates discovered in real-world deployment.

\section{Conclusion}
Identifying duplicates in ancient manuscripts is an important real-world problem. For AI-enabled Oracle Bone duplicates discovery, we integrate  unsupervised low-level feature matching with high-level character-based visual content matching for accurately and efficiently identifying the correct OB duplicates. We have discovered over 60 pairs of new OB duplicates in real-world deployment. In future work,  we will jointly use the dual modalities of rubbing and manual copies to conduct multi-modal OB duplicates discovery. We will also utilize our framework to discover  duplicates in other ancient manuscripts, including Bamboo slips, Turfan Manuscripts, Dead Sea Scrolls, etc.



\bibliography{ref}

\appendix

\appendix



\end{document}